\documentclass[twocolumn,english,prl,showpacs]{revtex4}
\usepackage[T1]{fontenc}
\usepackage[latin9]{inputenc}
\usepackage{amssymb}

\makeatletter

\def\journal #1, #2, #3, 1#4#5#6{{\sl #1~}{\bf #2}, #3 (1#4#5#6) }

\usepackage{mathrsfs}
\usepackage{float}
\usepackage{indentfirst}

\def\eqa{\begin{eqnarray}}
\def\eea{\end{eqnarray}}
\newcommand{\eq}{\begin{equation}}
\newcommand{\ee}{\end{equation}}

\makeatother

\usepackage{babel}

\usepackage{amsmath}
\usepackage{graphicx}

\begin{document}
\title{Antiferromagnetism of Repulsively Interacting Fermions in a harmonic trap}

\author{Jianqing Qi, Lei Wang, Xi Dai}

\affiliation{Beijing National Laboratory for Condensed Matter
Physics, and Institute of Physics, Chinese Academy of Sciences,
Beijing 100190, China}

\date{\today}

\begin{abstract}

We propose a Real-Space Gutzwiller variational approach and apply it
to a system of repulsively interacting ultracold fermions with spin
$\frac{1}{2}$ trapped in an optical lattice with a harmonic
confinement. Using the Real-Space Gutzwiller variational approach,
we find that in system with balanced spin-mixtures on a square
lattice, antiferromagnetism either appears in a checkerboard pattern
or forms a ring and antiferromagnetic order is stable in the regions
where the particle density is close to one, which is consistent with
the recent results obtained by the Real-Space Dynamical Mean-field
Theory approach. We also investigate the imbalanced case and find
that antiferromagnetic order is suppressed there.

\end{abstract}


\maketitle

\section{Introduction}

Ultracold atomic gases have attracted much attention {\cite{ATOM_1}}
since the first realization of Bose-Einstein condensation
{\cite{ATOM_2}}. In recent years, ultracold atoms in optical
lattices have stimulated a new wave of studying the many-body
problems. One can obtain optical lattices by confining the ultracold
atoms in periodic trapping potentials created with
counter-propagating laser beams {\cite{OPT_1}}. Owing to the large
degree of control over the optical lattice parameters such as the
geometry and depth of the potential, optical lattices provide an
ideal playground for studying fundamental condensed-matter physics
problems. Many remarkable phenomena, like the quantum phase
transition from a superfluid to a Mott-insulator in a Bose-Einstein
condensate with repulsive interaction {\cite{OPT_2}} and the
superexchange interactions with ultracold atoms {\cite{OPT_3}} have
been observed experimentally in optical lattices. In addition,
loading ultracold fermions as well as mixtures of bosonic and
fermionic quantum gases in optical lattices has also become a topic
of intensive study {\cite{OPT_4, OPT_5, OPT_6}}.

Although optical lattices have been providing an ideal stage for
both theoretical and experimental studies of fundamental problems in
condensed matter physics, when compared to true solid state system,
defects arise. For example, in optical lattices, an additional
harmonic confinement is always present due to the gaussian profile
of the laser beams {\cite{OPT_1}}. Although this harmonic
confinement is usually weak and varies slowly (typically around
10-200 Hz oscillation frequencies) compared to the confinement of
the atoms on each lattice site (typically around 10-40 kHz), it
generally leads to an inhomogeneous environment for the trapped
atoms. Therefore, in order to make problems more relevant to
condensed matter systems, investigating how the harmonic confinement
affects the behavior of atoms trapped in optical lattices is
important. Motivated by this, we take the ultracold fermions with
spin $\frac{1}{2}$ into consideration and concentrate on the
magnetic behavior of these particles in such a harmonic confinement.

For simplicity, in this paper we consider the square lattice with a
single orbital per site as a model, which can be described by the
famous Hubbard Hamiltonian. Hubbard model has been studied by
various methods such as variational Monte-Carlo method {\cite{VMC}}
and dynamical mean-filed theory {\cite{DMFT}}. Here we apply the
Gutzwiller approximation {\cite{GV_5}}, which was introduced by
Gutzwiller along with his proposal of Gutzwiller wave function
(GWF). It turns out that Gutzwiller approximation is exact in the
limit of infinite dimensions. Extensions to multi-band correlated
systems using Gutzwiller approximation were carried out by J. B\"{u%
}nemann \textit{et al.} {\cite{GV_6}}. Meanwhile, Gutzwiller
approximation was proved to be equivalent to slave-boson theories
{\cite{SLBT_1, SLBT_2, SLBT_3}} on a mean-field level for both
one-band case  {\cite{SLBT_4}} and multi-band case {\cite{SLBT_5,
SLBT_6}}. Gutzwiller approximation is usually used in homogenous
environment, here we extends it to inhomogeneous environment and
address the problem in real space. The organization of the paper is
as follows: first, we introduce the Hubbard Hamiltonian as well as
the Gutzwiller variational approach (GVA). Then we show how the
harmonic confinement potential and the repulsive interaction affect
the magnetism of the system in the case of balanced spin-mixtures
and then we present the results obtained in a imbalanced case.
Finally, we make some discussions and conclusions.

\section{Model and Method}

We apply the Hubbard model for repulsively interacting fermions in an
optical lattice. The Hamiltonian is described as
\begin{eqnarray}
 H & = & H_0+H_{int} \nonumber
\label{ham}
\end{eqnarray}
\begin{eqnarray}
 H_0 & = & -\sum_{\langle
 ij\rangle,\sigma}t_{ij}c_{i\sigma}^{\dagger}c_{j\sigma} \nonumber
\label{H0}
\end{eqnarray}
\begin{eqnarray}
H_{int} & = &
U\sum_{i}n_{i\uparrow}n_{i\downarrow}+\sum_{i\sigma}(V_{i}-\mu)
n_{i\sigma} \label{Hint}
\end{eqnarray}where $n_{i\sigma}=c^\dagger_{i\sigma}c_{i\sigma}$, and $c_{i\sigma}
(c^\dagger_{i\sigma}$) are fermionic annihilation (creation)
operators for an atom at the $i$th site with spin $\sigma$. $t_{ij}$
describes the hopping amplitude between nearest neighbor sites
$\langle ij\rangle$. If $i$ and $j$ are nearest neighbors,
$t_{ij}=t$, otherwise, $t_{ij}=0$. $U>0$ is the repulsive
interaction, $\mu_\sigma$ is the chemical potential and
$V_i=\frac{1}{2}\Omega^2r_i^2=V_0r_i^2$ is the external trapping
potential, in which $r_i$ is the distance measured from the center
of the system. As pointed out in reference {\cite{OPT_1}}, $\Omega$
is usually much smaller than the characteristic frequency of the
optical lattice, providing a spatially slowly varying chemical
potential.

Many methods, such as Hartree-Fock theory {\cite{HF}} and Real-Space
Dynamical Mean-Field Theory (R-DMFT) approach {\cite{RDMFT}}, have
been used to study the ground state of Hubbard model with a
confinement potential. Among these methods, R-DMFT approach is the
most accurate and reliable one, because it includes all the local
quantum fluctuation. However, the solution of Anderson impurity
model in each iteration step makes it very time-consuming. Here we
apply the Real-Space Gutzwiller variational approach (R-GVA) for
this model. We will show that the results obtained by R-GVA is
consistent with those obtained by R-DMFT approach.

The GVA has been proved to be quite efficient and accurate
{\cite{GV_1, GV_2, GV_3}} for the ground state studies of many
important phenomena in strongly correlated system, i.e. the Mott
transition, ferromagnetism and superconductivity{\cite{qianghua,qianghua2}}. It has also been
demonstrated {\cite{GV_4}} that GVA is as accurate as DMFT method
for the ground state properties, but much computationally cheaper,
which grants this approach much validity.

We first give a description of GVA for the ground state of
correlated electron model systems. There are $2$ different
spins and each of them could be either empty or
occupied, thus totally we have $2^2=4$ number of local configurations
$|\Gamma\rangle$ on a single site. Those possible configurations should not be
equally weighted, because electrons tend to occupy configurations
which have relatively lower energy. For this purpose, we could
construct projectors which can reduce the specified high energy configurations
$|\Gamma\rangle$ on site $i$
\begin{equation}
\hat{m}_{i \Gamma }=\left\vert{i,\Gamma} \right\rangle \left\langle{
i,\Gamma} \right\vert   \label{eq:m_I_projection}
\end{equation}%
which fufills,\begin{equation} \sum_{\Gamma }\hat{m}_{i\Gamma }=1
\label{eq:local_completeness}
\end{equation}%
since all the configurations ${|\Gamma\rangle}$ form a locally
complete set of basis.

In Eq.(\ref{ham}), if the interactions are absent, the ground state
is exactly given by the Hartree uncorrelated wave function (HWF)
$|\Psi _{0}\rangle $, which is a single determinant of single
particle wave functions. However, after turning on the interaction
terms, the HWF is no-longer a good approximation, since it contains
many energetically unfavorable configurations. In order to describe
the ground state better, the weights of those unfavorable
configurations should be suppressed. This is the main spirit of
Gutzwiller wave function (GWF). GWF $|\Psi _{G}\rangle $ is
constructed by acting a many-particle projection operator on the
uncorrelated HWF,
\begin{eqnarray}
&|\Psi _{G}\rangle =\mathcal{\hat{P}}|\Psi _{0}\rangle \nonumber \\
\mathcal{\hat{P}}& = \prod\limits_{i}\hat{P}_{i} =
\prod\limits_{i}\sum_{\Gamma}\lambda _{i\Gamma }\hat{m}_{i \Gamma}
\end{eqnarray}%

The projection operator $\mathcal{\hat{P}}$ is used to adjust the
weight of site configurations through parameters $\lambda _{i\Gamma
}$ ($0\leq \lambda _{i\Gamma }\leq 1$). The GWF falls back to
uncorrelated HWF if all $\lambda _{i\Gamma }=1$. On the other hand,
if $\lambda _{i\Gamma }=0$, the configuration $\Gamma $ of site $i$
will be totally removed. In this way, both the itinerant behavior of
uncorrelated wave functions and localized behavior of atomic
configurations can be described consistently, and the GWF will give
a more reasonable physical picture of correlated systems than HWF
does.

The evaluation of GWF is a difficult task due to its
multi-configuration nature. There are lots of efforts in the
literature, and the most famous one is Gutzwiller approximation
{\cite{GV_5}}. In this approximation, the inter-site correlation
effect has been neglected and the physics meaning was discussed in
{\cite{GV_1}} and {\cite{GV_7}}. The exact evaluation of the
single-band GWF in one dimension {\cite{one_dim}} and in the limit
of infinite dimensions {\cite{in_dim}} were carried out. It turns
out that Gutzwiller approximation is exact in the latter case.

The expectation value of Hamiltonian Eq.(1) is:
\begin{equation}
\langle H\rangle _{G}=\frac{\langle \Psi _{G}|H|\Psi _{G}\rangle
}{\langle
\Psi _{G}|\Psi _{G}\rangle }=\frac{\langle \Psi _{0}|\mathcal{\hat{P}}H%
\mathcal{\hat{P}}|\Psi _{0}\rangle }{\langle \Psi _{0}|\mathcal{\hat{P}}%
^{2}|\Psi _{0}\rangle }  \label{eq:H_eff_gut}
\end{equation}%
We note that by choosing $\lambda _{i\Gamma }=\sqrt{\frac{m_{i\Gamma }}{%
m_{i\Gamma }^{0}}}$, $|\Psi _{G}\rangle $ is normalized under GA.
 $\langle
\Psi _{G}|\Psi _{G}\rangle =\prod\limits_{i}\langle \Psi _{0}|\hat{P}%
_{i}^{2}|\Psi _{0}\rangle =\prod\limits_{i}\sum_{\Gamma }\frac{m_{i\Gamma }}{%
m_{i\Gamma }^{0}}\langle \Psi _{0}|\hat{m}_{i;\Gamma }|\Psi
_{0}\rangle =\prod\limits_{i}\left( \sum_{\Gamma }m_{i\Gamma
}\right) =1$. Here $m_{i\Gamma}$ is the weight of configuration
$\Gamma _{i}$, $m_{i\Gamma }=\langle \Psi _{G}|\hat{m}_{i\Gamma
}|\Psi _{G}\rangle$ and $m_{i\Gamma }^{0}=\langle \Psi
_{0}|\hat{m}_{i\Gamma }|\Psi _{0}\rangle$. In the first equality we
separate the average of a projection operator string into the
product of single site averages.

The expectation value of kinetic energy is
\begin{align}
& \langle \Psi _{G}|H_{0}|\Psi _{G}\rangle \notag\\
=& \sum_{i,j,\sigma }z_{i\sigma }z_{j\sigma } t_{ij}\langle \Psi
_{0}|c_{i\sigma }^{\dagger }c_{j\sigma }|\Psi _{0}\rangle
\label{kin}
\end{align} where
\begin{equation}  \label{eq:zfac}
z_{i\sigma }=\sum_{\Gamma_i,\Gamma_i^{\prime}}\frac{\sqrt{m_{\Gamma
_{i}}m_{\Gamma _{i}^{\prime }}}D_{\Gamma _{i}^{\prime}\Gamma _{i}}^{\sigma }%
}{\sqrt{n_{i\sigma }^{0}\left( 1-n_{i\sigma }^{0}\right) }}
\end{equation}
$\bigskip $ with
 $D_{\Gamma ^{\prime }\Gamma }^{\sigma }=|<\Gamma
^{\prime }|c_{i\sigma }^{\dagger }|\Gamma >|$, $0\leq z_{i\sigma
}\leq 1$.

while for the interaction part of the Hamiltonian
\begin{align}
& \langle \Psi _{G}|H_{int}|\Psi _{G}\rangle  \notag \\
& =\sum_{i}\sum_{\Gamma }E_{\Gamma }\frac{m_{i\Gamma }}{m_{i\Gamma
}^{0}}
\langle \Psi _{0}|\hat{m}_{i\Gamma }|\Psi _{0}\rangle  \notag \\
& =\sum_{i}\sum_{\Gamma }E_{\Gamma }m_{i\Gamma }  \label{int}
\end{align}

Putting Eq.(\ref{kin}) and Eq.(\ref{int}) together, we have the
following equation for the limit of infinite dimensions
\begin{align}
& \langle H\rangle _{G} \notag \\
& =\sum_{i\neq j, \sigma }t_{i,j}z_{i,\sigma }z_{j,\sigma }\langle
c_{i\sigma }^{\dagger }c_{j\sigma }\rangle_{0}+\sum_{i,\sigma }\epsilon _{i\sigma }n_{i,\sigma }^{0} +\sum_{i,\Gamma }E_{i\Gamma }m_{i\Gamma } \notag\\
& =\langle 0|H_{eff}|0\rangle
\label{eq:mgutzH}
\end{align}
In an inhomogeneous systems, the spatial dependence of $z_{i\sigma
}$ is preserved and the variation is in the $4N_{s}$ parameter
space, where $N_{s}$ is the number of sites. We adopt the following
algorithm to minimize $\langle H\rangle _{G}$. We begin with solving
$H_{eff}$ where the $Z$-factors are fixed, from which we compute the
expectation value of the Fermionic operators on the ground state.
Then the minimization of the Gutzwiller variational parameter $m$ is
done in the alternating least squares (ALS) scheme, in which we fix
the $m_{\Gamma}$s on all but the current site and the problem
reduces to quadratic optimization which is solved as an
eigen-problem. Using Eq.(\ref{eq:zfac}), one could compute the
$Z$-factor on each site, and then they are plugged into the
non-interacting model $H_{eff}$ as parameters. The iteration is
finished when the difference of $Z$-factors from two step is less
then the given precision, say $10^{-6}$. In general, there is no
guarantee that the ALS method will converge to the global optimum
and the convergence of the iteration. However, in practice, this
does not seem to occur as long as one varies the parameters in the
Hamiltonian adiabatically.

In the following part, we consider this model on a  $(24\times24)$
square lattice at half filling (one particle per site) and set $t$ as the unit of energy.

\section{Results and Discussions}

\begin{figure}
\includegraphics[clip,scale=0.2]{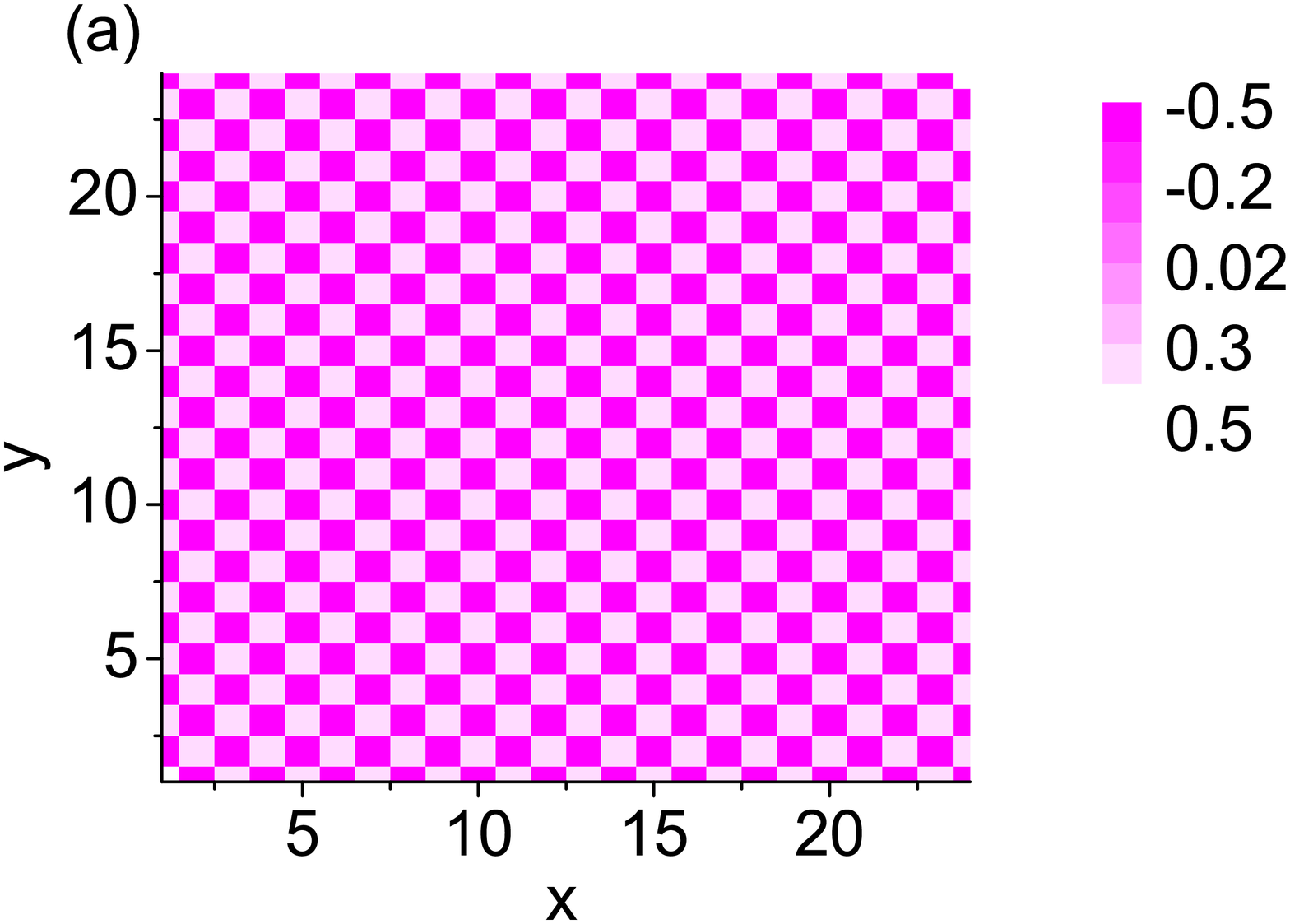}
\includegraphics[clip,scale=0.2]{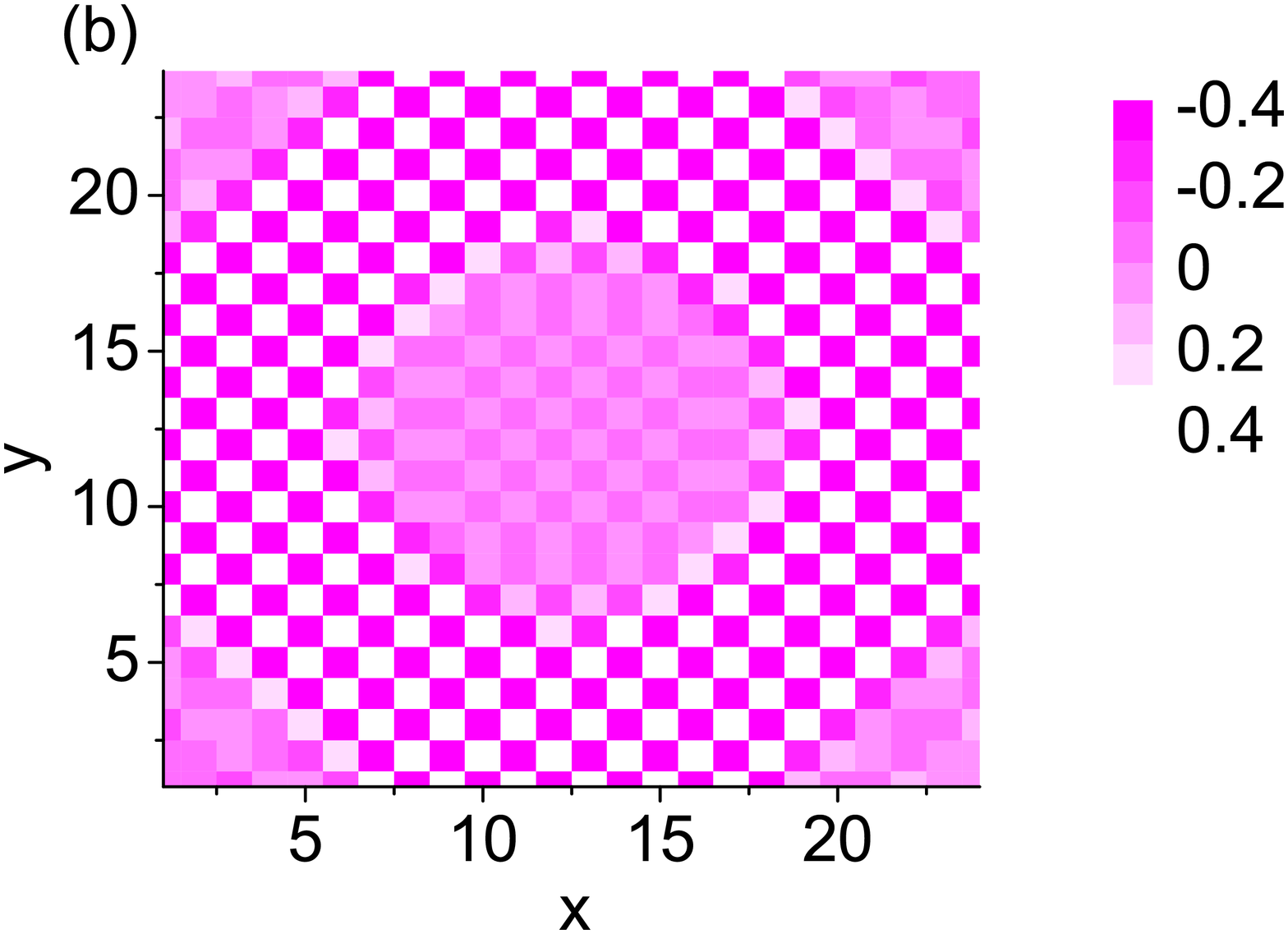}
\includegraphics[clip,scale=0.2]{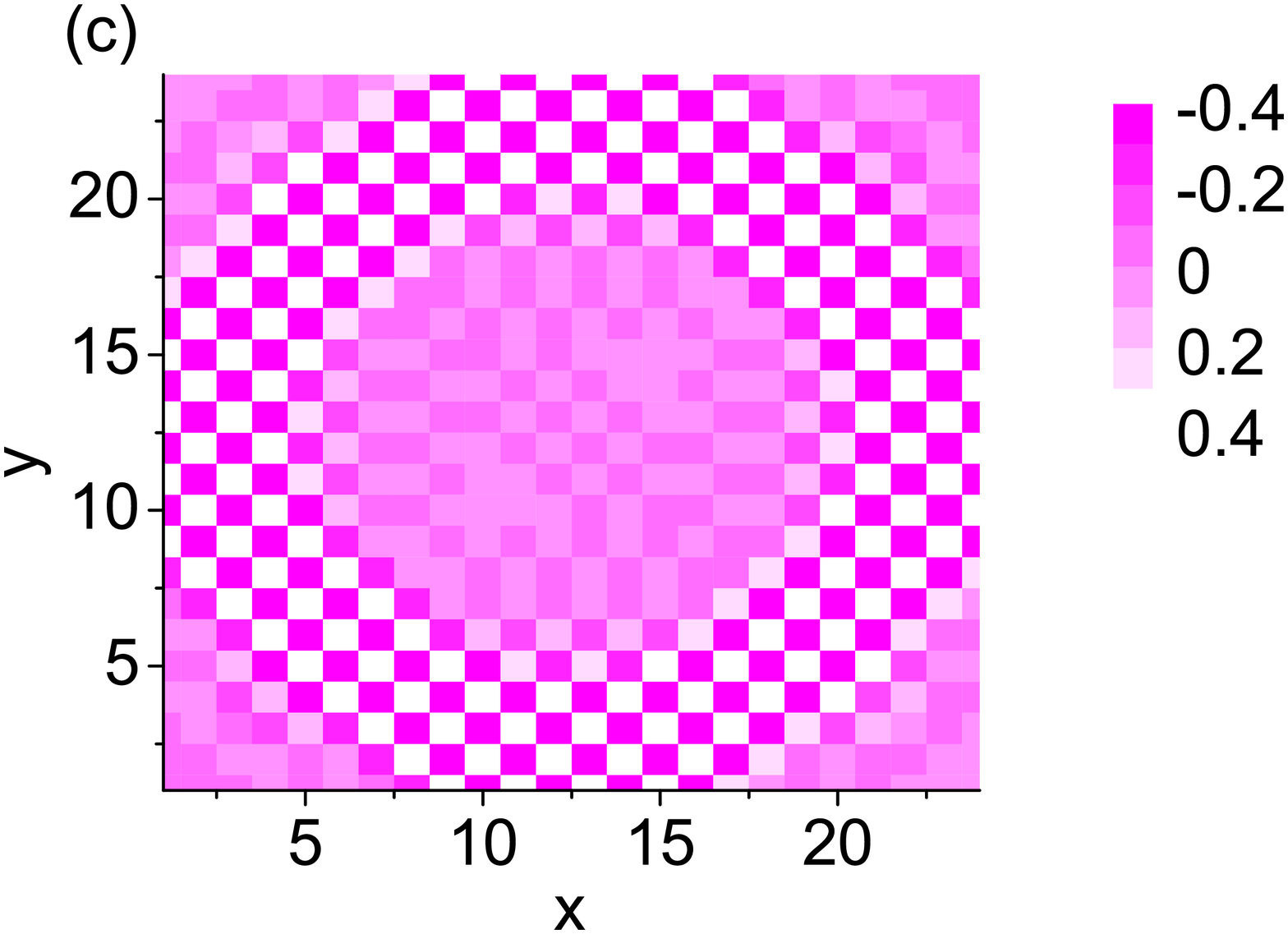}
\includegraphics[clip,scale=0.2]{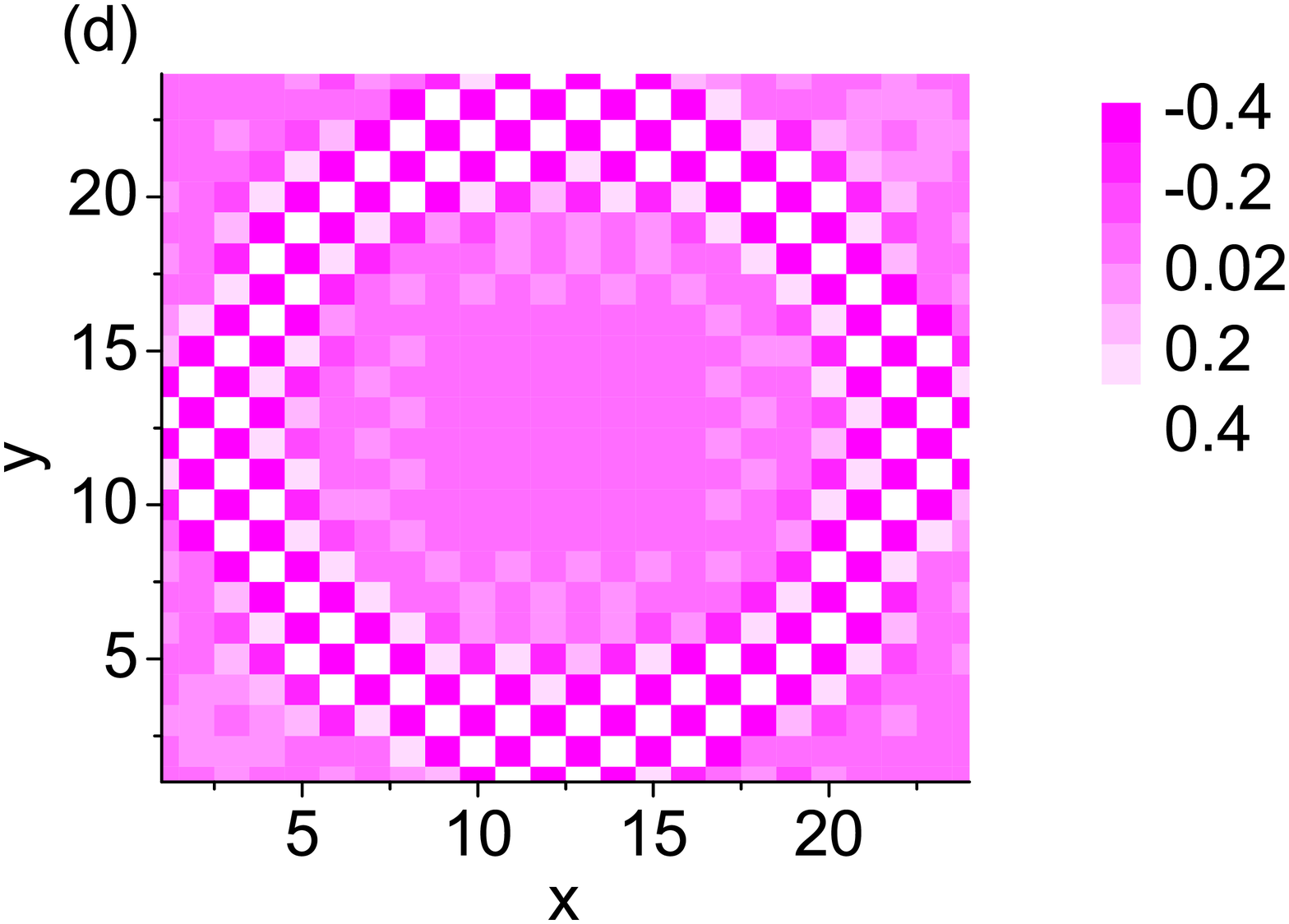}
\caption{(Color online) Real-Space magnetization profiles for $U=5$
on square lattice $(24\times24)$ at half filling, when (a) $V=0.01$,
(b) $V=0.02$, (c) $V=0.03$, (d) $V=0.04$. AF region shrinks as the
confining potential increases.} \label{V_pattern}
\end{figure}

Now we present the numerical results obtained with the R-GVA. We
focus on the spatial dependence of magnetization and particle
distribution at different parameters. We first consider the balanced
situation in which the number of spin-$\uparrow$ particles is the
same to that of the spin-$\downarrow$ ones. We begin with the discussion on
the effect of the harmonic confinement $V$. First we fix the
repulsive interaction $U=5$. The spatial distribution of
magnetization at different strengths of $V$ is shown in Fig.
\ref{V_pattern}.

We find that antiferromagnetic(AF) order exists even with the
presence of the inhomogeneous harmonic confinement. It is seen
clearly from Fig. \ref{V_pattern} that how the pattern of
magnetization evolves as the harmonic confinement $V$ increases. As
the confinement potential $V$ is enhanced, the antiferromagnetism
changes from a uniform checkerboard structure to a ring, where the
filling is close to $1$.

\begin{figure}
\includegraphics[clip,scale=0.25]{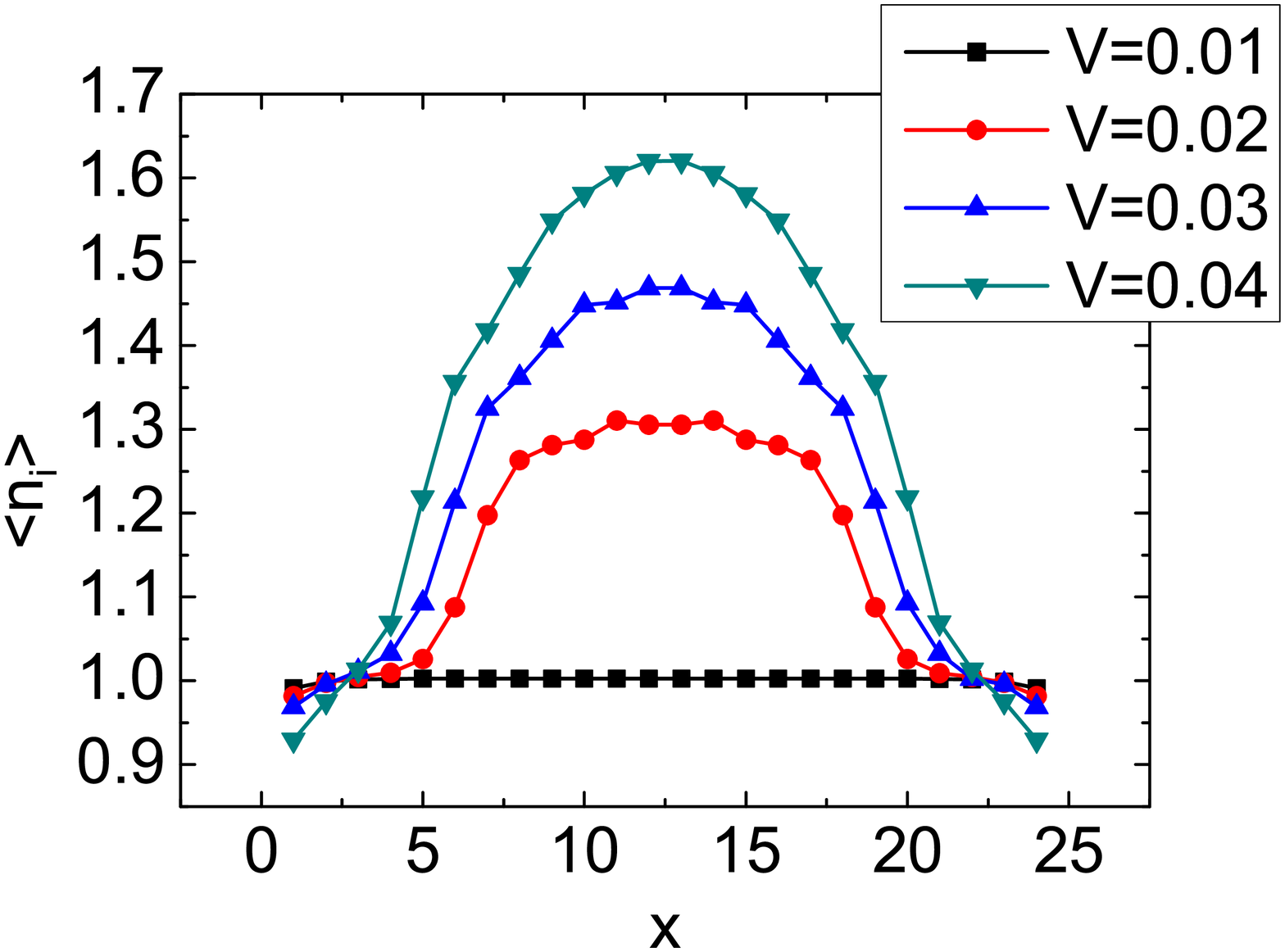}
\includegraphics[clip,scale=0.25]{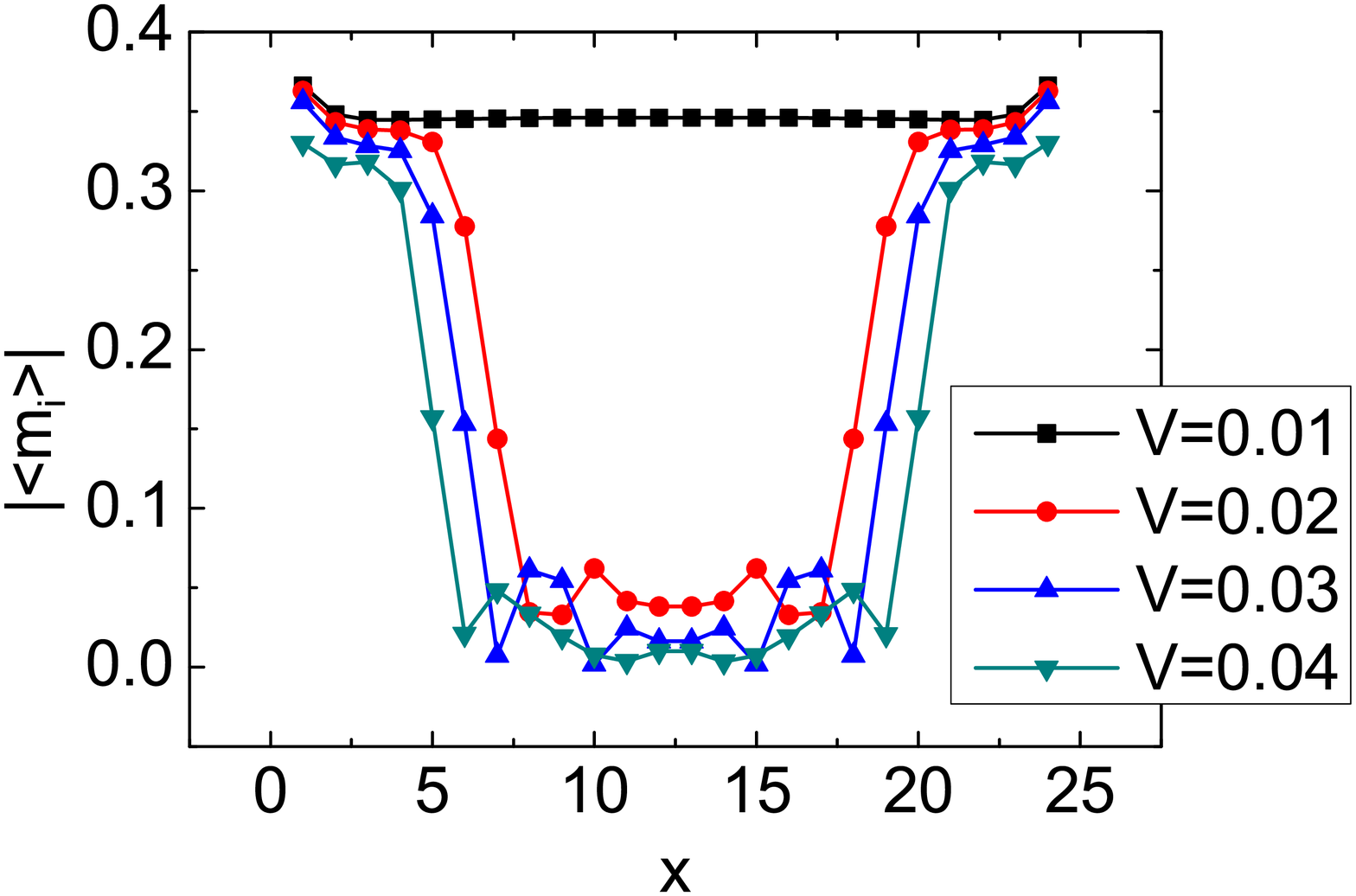}
\caption{(Color online) Particle density and the absolute value of
staggered magnetization as the function of the distance along $y=12$
for $U=5$ at different confining potentials at half filling: (a) the
density profile; (b) the staggered magnetization.} \label{V_cross}
\end{figure}

To make the problem more explicit, we also get the particle and spin
density profiles along x-direction. In Fig. \ref{V_cross} (a) and
(b), we present the local density $\langle n\rangle_i$ and the
absolute value of the staggered magnetism $\langle m_i \rangle$ as
the function of the distance along $y=12$, where $\langle n_i
\rangle=\langle n_{i\uparrow}\rangle+\langle n_{i\downarrow}\rangle$
and $\langle m_i \rangle =\frac{1}{2}(\langle n_{i\uparrow}
\rangle-\langle n_{i\downarrow\rangle})$. We find that in the
presence of the confinement potential $V$, the antiferromagnetic phase is
stable when the local density is close to $1$, which is consistent
with the result obtained in reference {\cite{RDMFT}}. The results
obtained here confirm the role that the harmonic confinement plays
in affecting the antiferromagnetic pattern among the fermions in
optical lattices. As pointed out in {\cite{RDMFT}}, these results
are important for the ongoing attempt to realize antiferromagnetic
state of fermions with repulsive interactions in periodic
potentials.

\begin{figure}
\includegraphics[clip,scale=0.2]{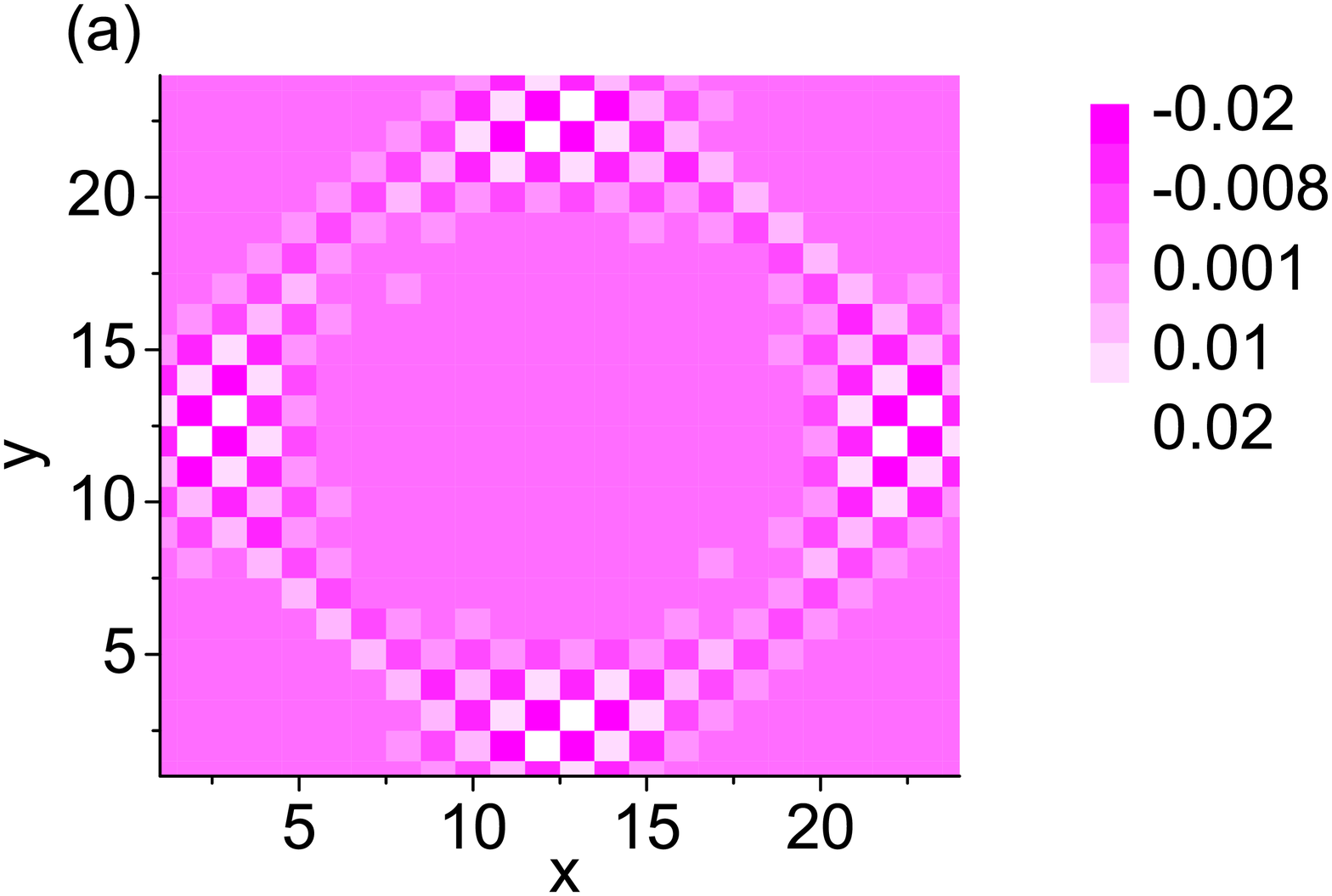}
\includegraphics[clip,scale=0.2]{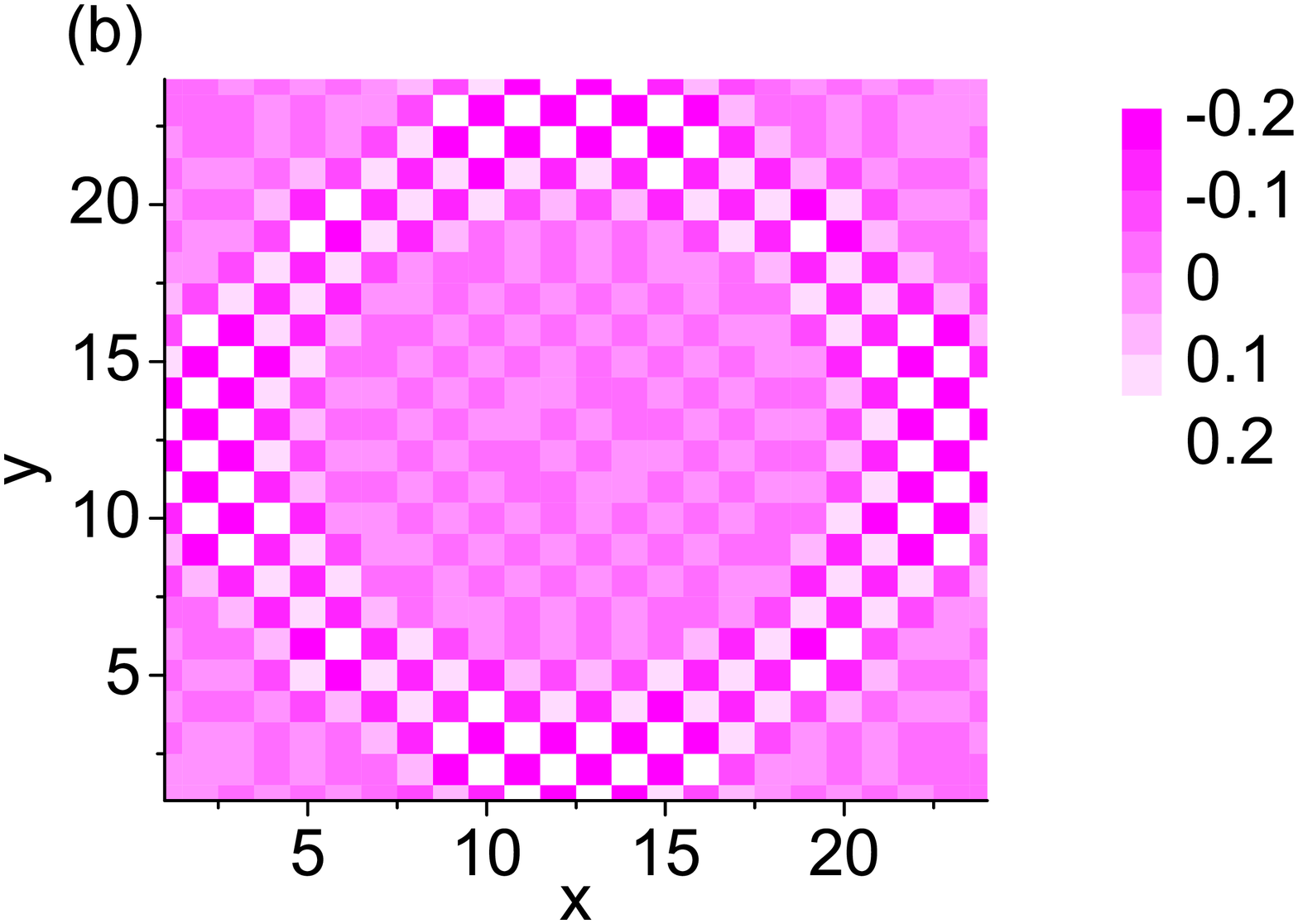}
\includegraphics[clip,scale=0.2]{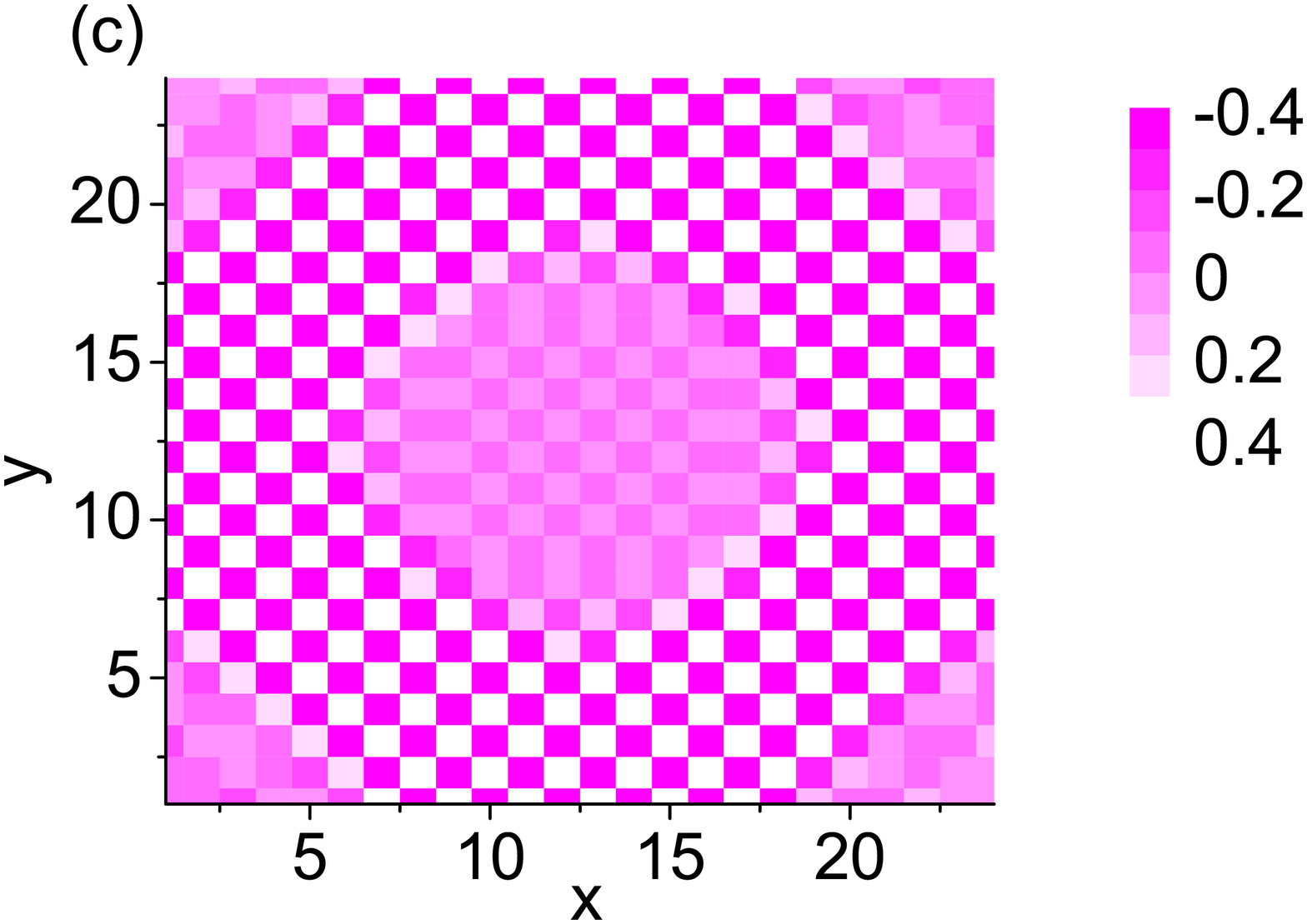}
\includegraphics[clip,scale=0.2]{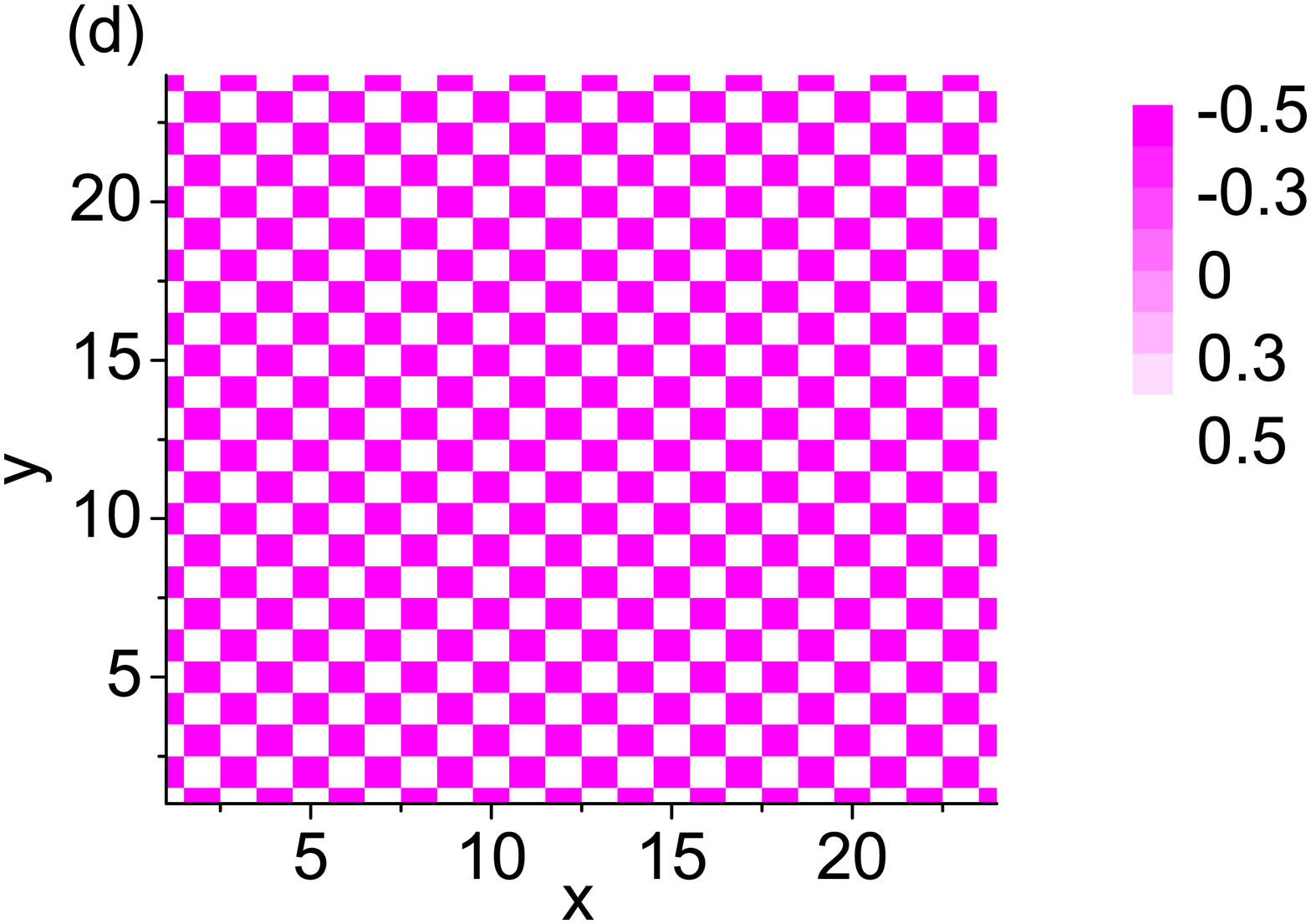}
\caption{(Color online) Real-Space magnetization profiles for
$V=0.02$ on square lattice $(24\times24)$ at half filling, when (a)
$U=2$, (b) $U=3$, (c) $U=5$, (d) $U=9$. The Af region expands as $U$
increases.} \label{U_pattern}
\end{figure}

\begin{figure}
\includegraphics[clip,scale=0.25]{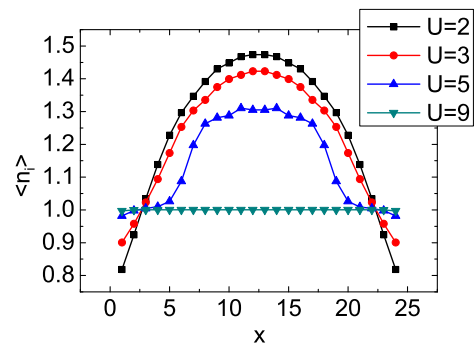}
\caption{(Color online) Particle density profile along $y=12$ for
$V=0.02$ at different strengths of repulsive interaction $U$ at half
filling. The particle density is more and more close to $1$ as $U$
increases.} \label{U_cross}
\end{figure}

Next, we concentrate on the effect of the repulsive interaction $U$.
Experimentally, $U$ could be tuned by the Feshbach resonance
technique. We first set the confining potential $V$ as 0.02. The
spatial dependence of magnetization and the local particle
distribution for at different strengths of $U$ are presented in Fig.
\ref{U_pattern} and Fig. \ref{U_cross}. We know that the ground
state of ultra-cold fermions loading in an optical lattice without
trap follows the spin density wave (SDW) mean filed prediction at
weak coupling. Approaching the strong coupling limit, the large
repulsive interaction drives the system to an AF insulator phase.
From Fig. \ref{U_pattern} and Fig. \ref{U_cross}, we can see that
the confining potential $V$ plays a dominant role at weak coupling
and the SDW state is suppressed, while at strong coupling, the
repulsive interaction $U$ plays a dominant role and the AF order is
stable.

\begin{figure}
\includegraphics[clip,scale=0.2]{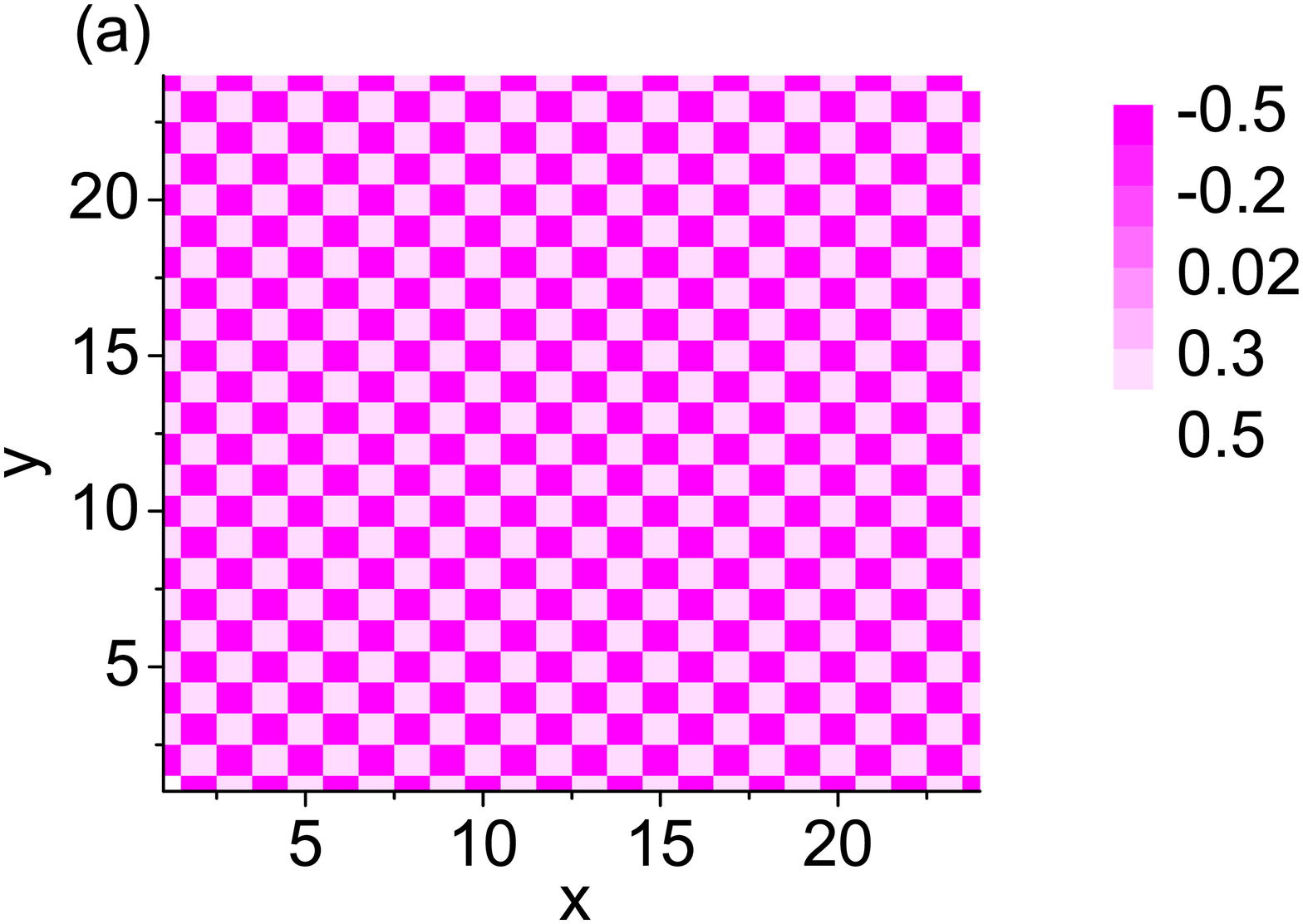}
\includegraphics[clip,scale=0.2]{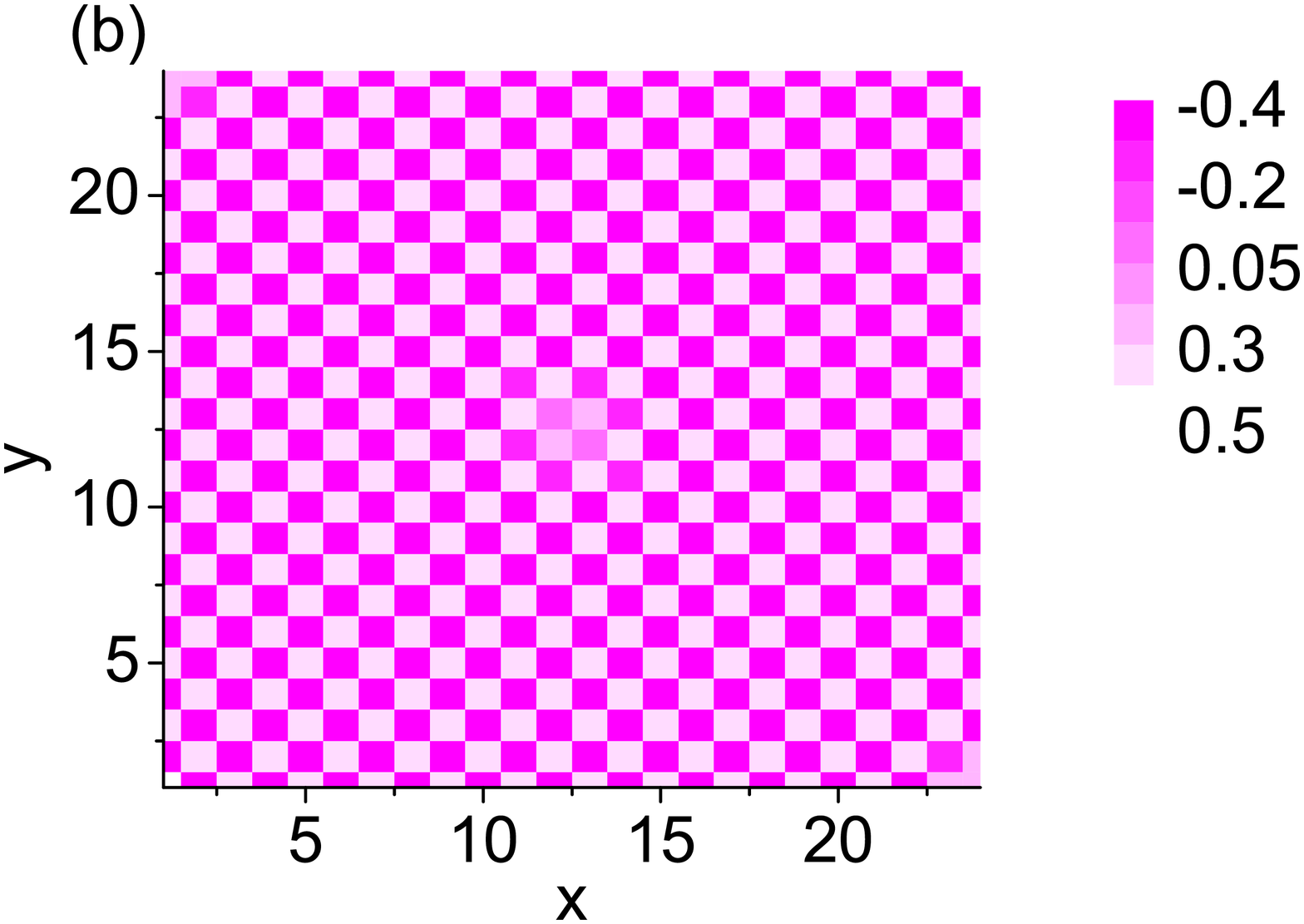}
\includegraphics[clip,scale=0.2]{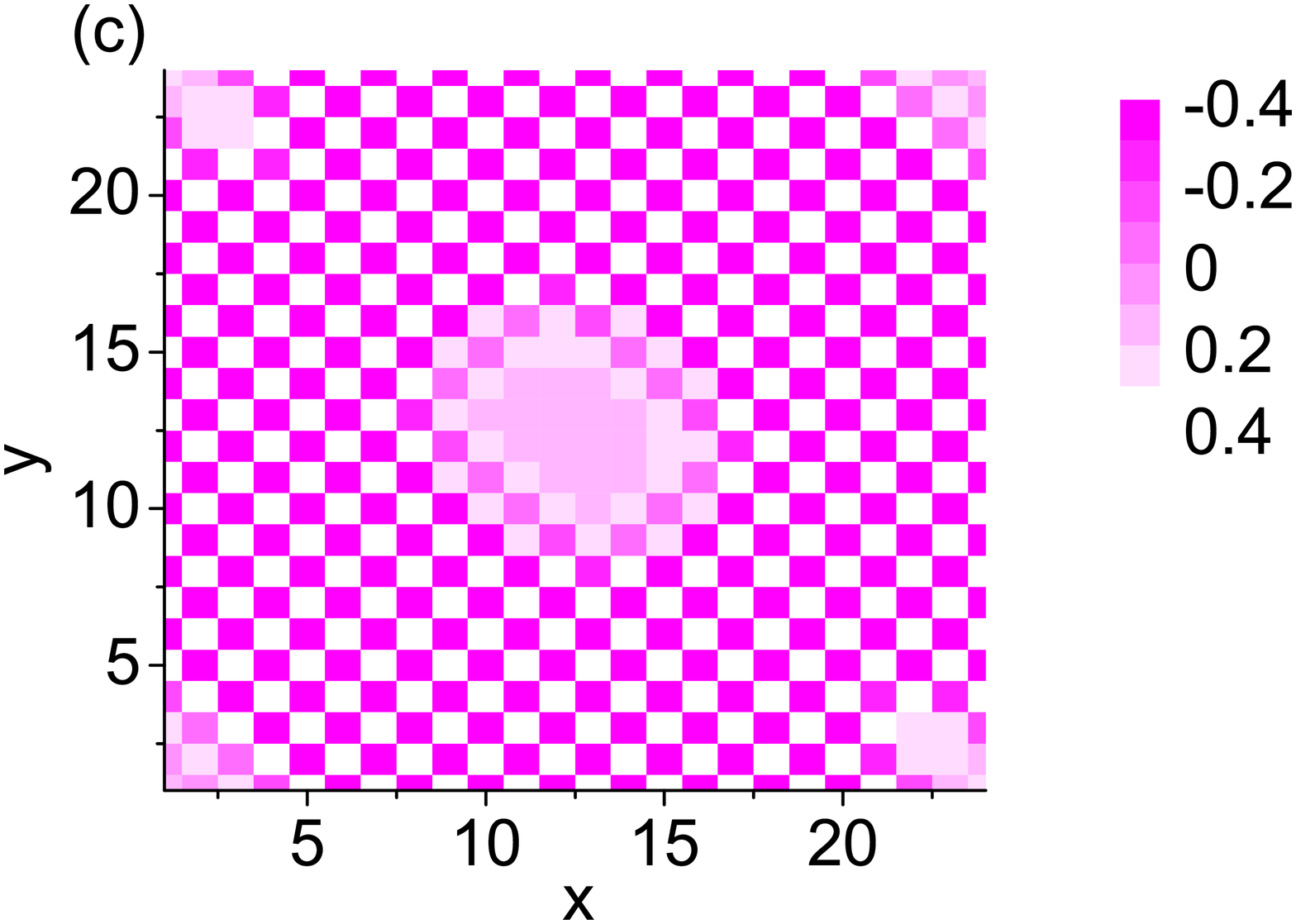}
\includegraphics[clip,scale=0.2]{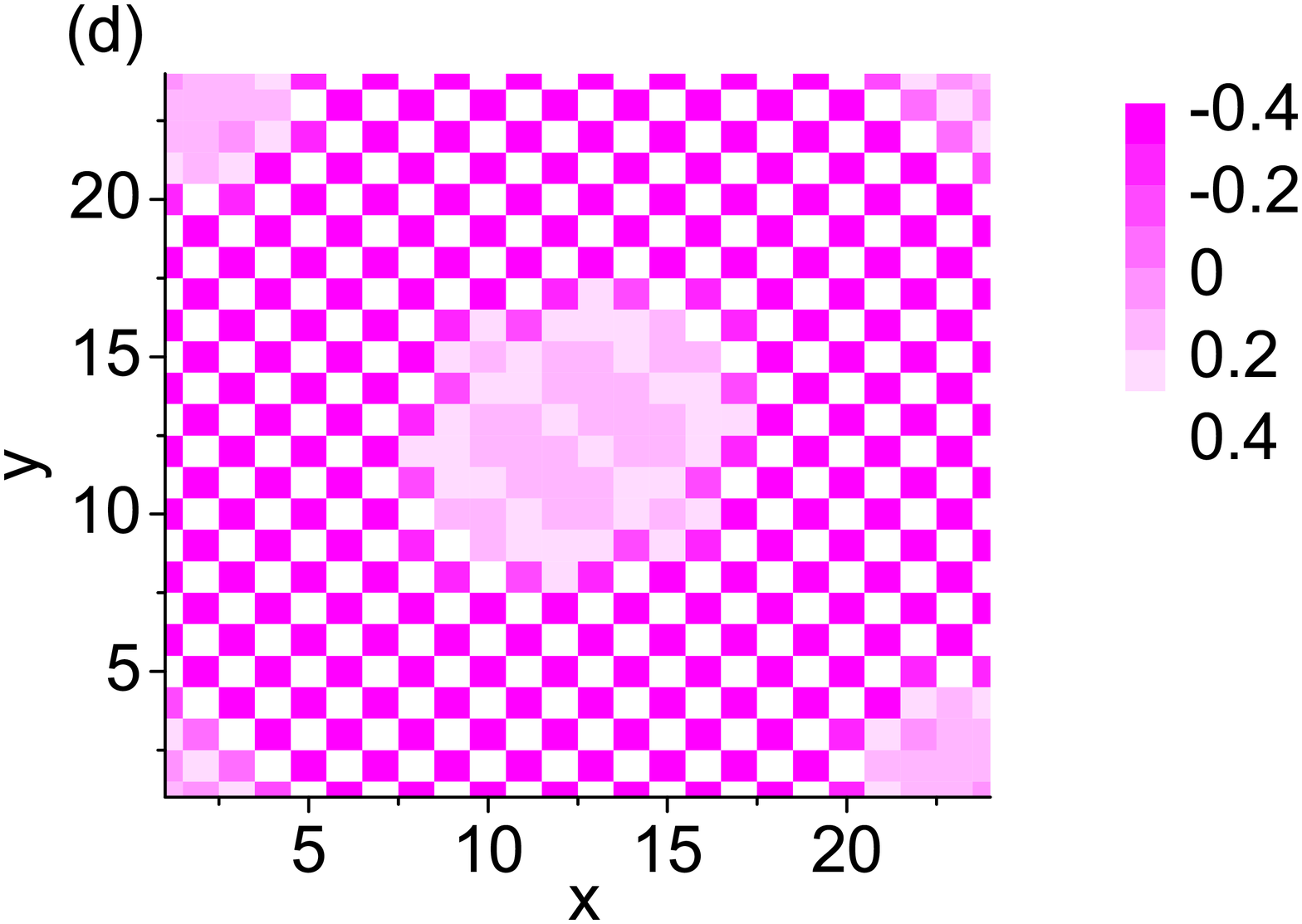}
\caption{(Color online) Real-Space magnetization profiles for
$V=0.01$ and $U=5$ on square lattice $(24\times24)$ for imbalanced
spin-mixtures at half filling, when (a) $N_\downarrow =288,
N_\uparrow =288$; (b) $N_\downarrow =286, N_\uparrow =290$; (c)
$N_\downarrow =275, N_\uparrow =301$; (d) $N_\downarrow =270,
N_\uparrow =306$. The AF order decreases as the imbalance is
enhanced.} \label{im_pattern}
\end{figure}

\begin{figure}
\includegraphics[clip,scale=0.25]{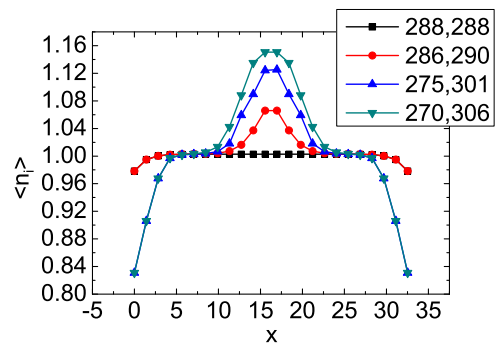}
\caption{(Color online) Particle density profile as the function of
the distance along $y=x$ for $U=5$ and $V=0.01$ for different
spin-mixtures at half filling. The particle density in the center of
the system increases as the imbalance is enhanced.} \label{im_cross}
\end{figure}

We now investigate the case of imbalanced spin-mixtures, i.e. when
$N_{\uparrow} \not= N_{\downarrow}$. The spatial dependence of
magnetism and the particle density of sublattice at different
strengths of imbalance are presented in Fig. \ref{im_pattern} and
Fig. \ref{im_cross}. We find that as the imbalance is enhanced, the
AF order decreases. In balanced system, antiferromagnetism competes
with the confining potential $V$. Upon imbalanced spin-mixtures, it
follows that an equivalent magnetic field is added into the system,
therefore the AF order is destroyed.

\section{Experimental Signatures}
Spatial distribution of spin density in a harmonic trap predicted
in this paper could be detected by Bragg scattering {\cite{MET_1}},
and by spatial microwave transition and spin-changing collisions
techniques, which measure the integrated density profiles along
chosen directions {\cite{MET_2}}.

\section{Conclusion}
In conclusion, we have developed the fast Real-Space Gutzwiller variational
approach which is suitable for the fast determination of the grounds
state phase diagram of the inhomogeneous strongly correlated
systems. With this method, we have studied both balanced and
imbalanced case of fermions with spin $\frac{1}{2}$ trapped in an
optical lattice with a harmonic confinement potential. We find that the trap
potential tends to destroy the AF order in the center as well as the
edge of the trap, leaving a ring of AF region with local density
close to $1$. The AF order is suppressed for imbalanced system.
These results are meaningful for the ongoing attempt to realize AF
in the optical lattices. We anticipate that this R-GVA scheme could
also be applied to other systems, such as a strongly interacting
Bose-Fermi mixture in a harmonic trap.

\end{document}